\documentclass[aps,prl,twocolumn,showpacs]{revtex4}
\usepackage{amsbsy,latexsym}
\usepackage{amsfonts}
\usepackage{amssymb}
\usepackage[mathscr]{eucal}

\newcommand{\al}{\alpha}
\newcommand{\be}{\beta}
\newcommand{\ga}{\gamma}
\newcommand{\de}{\delta}
\newcommand{\la}{\lambda}
\newcommand{\id}{\mathbb I}
\newcommand{\CF}{\mathsf n}
\newcommand{\aver}[1]{\langle #1 \rangle}
\newcommand{\tr}{{\rm tr}\,}

\newcommand{\ket}[1]{\vert #1 \rangle}
\newcommand{\bra}[1]{\langle #1 \vert}

\newcommand{\D}[3]{\mathfrak D^{(#1)}_{#2 #3}}
\newcommand{\Dc}[3]{\mathfrak D^{(#1) *}_{#2 #3}}

\newcommand{\U}[1]{U^{(#1)}}

\begin{document}
\title{Aligning Reference Frames Using
Quantum States}
\author{E.~Bagan, M.~Baig, and R.~Mu{\~n}oz-Tapia}
\affiliation{Grup de F{\'\i}sica Te{\`o}rica \& IFAE, Facultat de
Ci{\`e}ncies, Edifici Cn, Universitat Aut{\`o}noma de Barcelona, 08193
Bellaterra (Barcelona) Spain}
\date{\today}

\begin{abstract}
We analyze the problem of sending, in a single transmission, the
information required to specify an orthogonal trihedron or
reference frame through a quantum channel made out of $N$
elementary spins. We analytically obtain the optimal strategy,
i.e., the best encoding state and the best measurement. For large
$N$, we show that the average error goes to zero linearly in
$1/N$. Finally, we discus the construction of finite optimal
measurements.
\end{abstract}

\maketitle

Can a system of $N$ elementary spins be used to communicate in a
single transmission the orientation of three mutually orthogonal
unit vectors (orthogonal trihedron)? A positive answer would,
e.g., enable two distant parties (Alice and Bob) to establish a
common reference frame using just a quantum channel. This question
was addressed twenty years ago by Holevo~\cite{holevo} who
concluded that if such a quantum system has a well defined total
spin $J$ the best the sender (Alice) can attempt to achieve is to
transmit the orientation of {\em at most one} of the three
vectors. There has recently been renewed interest in this
simpler, more manageable, problem of sending a single direction,
and reformulations and extensions of the original question abound
in the literature~\cite{mp,derka,lpt,gp,massar,us,pp1,us2,us3}
(related issues can also be found in \cite{related}). In all the
cases, optimal communication involves collective (entangled)
measurements and an accurate choice of the messenger quantum
states.

 In this letter, we will be concerned with the more
complex problem of sending  the information that specifies an
orthogonal trihedron (OT). We will demonstrate that by encoding
the relevant geometrical information in a particular class of
states  one overcomes the limitations foreseen by Holevo and a
good transmission is possible. These states can be written as a
simple  superposition of states belonging to each of the
irreducible representations of $\mathrm{SU(2)}$ that appear in
the Hilbert space of the $N$ spins. They have maximal third
component of the total spin within each representation, i.e, in
standard notation are of the form $\sum C_j \ket{j,m=j}$
 (therefore they  are {\em not}
eigenstates of either $\vec J\,{}^2$ or $ J_z\,{}$). The quality
of the optimal communication strategy is shown to increase with
$N$ and in the limit $N\to\infty$ the average error, $\aver{h}$,
goes to zero. For large $N$ we obtain an analytical estimate of
this error, $\aver{h}\gtrapprox 8/N$. We would like to emphasize
that despite the apparent difficulty of the problem~\cite{pp2},
an analytical treatment is possible, which provides us with a
physical insight of the underlying quantum aspects involved in
the communication process.

Let us suppose Alice has a system of $N$ spins which she wishes
to use to tell Bob an OT, $\CF=\{\vec n_{1}, \vec n_{2},\vec
n_{3}\}$. By performing quantum measurements, Bob will be able to
reconstruct this OT with some accuracy and will make the guess
$\CF'=\{\vec n'_{1}, \vec n'_{2},\vec n'_{3}\}$. The obvious
parametrization of the different OTs is provided by the Euler
angles $\al$, $\be$, $\ga$, of the rotations that map
$\CF_{0}=\{\vec x,\vec y,\vec z\}$ into $\CF$, $\CF'$.  We will
use $g$ as a shorthand for the three Euler angles, i.e.,
$g=(\al,\be,\ga)$. Following Holevo~\cite{holevo}, we may
quantify the quality of the communication strategy by evaluating
the mean value of the error (or average error) defined for each
individual measurement by
\begin{equation}
    h(g,g')=\sum_{a=1}^3|\vec n_{a}-\vec n_{a}'|^2=
    \sum_{a=1}^3|\vec n_{a}(g)-\vec n_{a}(g')|^2.
    \label{ebc-6.5-6}
\end{equation}
Assuming the
OT are chosen from an isotropic distribution, and denoting by
$p_{g'}(g)$ the conditional probability of Bob guessing $\CF(g')$ if
Alice's OT is $\CF(g)$, one has
\begin{equation}
    \aver{h}=\int dg\int dg' h(g,g') p_{g'}(g),
    \label{ebc-6.5-7}
\end{equation}
where $dg$ is the Haar measure of the rotation group, ${\rm
SU(2)}$, which in terms of the Euler angles reads $dg=\sin\be\,
d\be d\al d\ga /8\pi^2$. Covariance implies
that~(\ref{ebc-6.5-7}) can be written as
\begin{eqnarray}
    &&\aver{h}=\int dg h(g,{\bf 0}) p_{{\bf 0}}(g),
    \label{ebc-6.5-8}
\end{eqnarray}
where $\bf0$ stands for $(\al,\be,\ga)=(0,0,0)$. One can easily check
that
\begin{equation}
    h(g,{\bf0})=6-2\tr \U{1}(g),
    \label{ebc.6.5-9}
\end{equation}
where $\U{j}$ is the $\rm SU(2)$ irreducible representation of
spin $j$, whose elements we write as
$\D{j}{m}{m'}(g)=\bra{j,m}\U{1}(g)\ket{j,m'}$. One also has $
t\equiv\tr \U{1}(g)=\sum_{m}
\D{1}{m}{m}(g)=\cos\be+(1+\cos\be)\cos(\al+\ga)$.
 We see that the
values of $t$ lay in the real interval $[-1,3]$. The value $t=3$
corresponds to perfect determination of Alice's OT and implies
that $\aver{h}=0$. Note also that $\aver{h}=6-2\aver{t}$. Random
guessing implies $\aver{t}=0$ ($\aver{h}=6$), while perfect determination of one
axis and random guessing of the remaining two imply $\aver{t}=1$
($\aver{h}=4$).

The most general quantum state Alice can use has the form
$\ket{A(g)}=U(g)\ket{A}$. Here $U(g)=\bigoplus_{j}U^{(j)}$ and
\begin{equation}
    \ket{A}=\sum_{j} \ket{A^j}=\sum_{j,m} A_{m}^j\ket{j,m};\qquad
\sum_{j,m}|A_{m}^j|^2=1,
    \label{ebc-6.5-1}
\end{equation}
where $j$ runs from $0$ to $N/2$ (for simplicity we will only
consider $N$ even) and $m$ runs from $-j$
to $j$. $\ket{A}$ is a fixed reference state associated with the OT $\CF_{0}$.

Likewise, we may write a reference state $\ket{B}$ from which we
can construct Bob's projectors of his Positive Operator Valued
Measurement (POVM) . The  general form of the state is
\begin{equation}
    \ket{B}=\sum_{j}\sqrt{2j+1}\ket{B^j};\quad
    \ket{B^j}=\sum_{m} B_{m}^j\ket{j,m},
    \label{ebc-6.5-2}
\end{equation}
where the square root is introduced for later convenience, and the
projectors are
\begin{equation}
    O(g)=U(g)\ket{B}\bra{B}U^\dagger(g).
    \label{ebc-6.5-3}
\end{equation}
We will first consider continuum POVMs for simplicity but finite ones
can also be constructed, as will be explained below. The condition
$\id=\int dg\,O(g)$ requires that
\begin{equation}
    \sum_{m=-j}^{j}|B_{m}^j|^2=1,\quad \forall j,
    \label{ebc-6.5-4}
\end{equation}
as can be easily shown with the help of the orthogonality
relations
\begin{equation}
    \int dg\, \D{J}{M}{m}(g)\Dc{J'}{M'}{m'}(g)={\de^{JJ'}\de_{MM'}\de_{mm'}
    \over 2J+1}.
    \label{ebc-6.5-5}
\end{equation}
Quantum Mechanics tells us that
$p_{{\bf0}}(g)=|\bra{B}U(g)\ket{A}|^2$, hence we have
\begin{equation}
    \aver{t}=\int dg\,|\bra{B}U(g)\ket{A}|^2\tr{\U{1}}.
    \label{ebc-6.5-11}
\end{equation}
In terms of the components of $\ket{A}$ and $\ket{B}$ the last
expression reads
\begin{equation}
 \aver{t}=\sum_{ljl'j'}\sum_{\mbox{\scriptsize$\begin{array}{c}mn
 \\[-.05cm]m'n'\end{array}$}} A^{l*}_{n}
 A^j_{m}B^{l}_{n'} B^{j*}_{m'} M^{lj}_{nmn'm'} ,
 \label{ebc-6.5-12}
\end{equation}
where
\begin{eqnarray}
 M^{lj}_{nmn'm'}&=&\sqrt{(2l+1)(2j+1)}\nonumber\\
 &\times&
 \int dg\,\tr\U{1}{}{}(g)\,\D{j}{m'}{m}(g)\Dc{l}{n'}{n}(g)
 \nonumber \\
&=&\sqrt{(2l+1)(2j+1)}\nonumber\\
 &\times&
 \sum_{M}\langle1Mjm|ln\rangle\langle1Mjm'|ln'\rangle,
 \label{ebc-6.5-13''}
\end{eqnarray}
and the last terms in brackets are the usual
Clebsch-Gordan coefficients.

The optimal strategy is the one that maximizes $\aver{t}$. It is
tempting  to introduce Lagrange multipliers $\la$ and $\mu^{j}$
for the normalization constrains (\ref{ebc-6.5-1})
and~(\ref{ebc-6.5-4}) respectively and follow the standard
maximization procedure. Analytical results along this line seem
hard to obtain~\cite{pp2}. We will, thus, try to develop a more
physical picture of Eqs.~\ref{ebc-6.5-11}--\ref{ebc-6.5-13''}
which will lead us to a
stunning  simplification of  the problem.

Notice that Eqs.~\ref{ebc-6.5-11}--\ref{ebc-6.5-13''}
can also be written in a compact
form  as
\begin{eqnarray}
    \aver{t}&=&\sum_{lj}{\sqrt{(2l+1)(2j+1)}\over3}
    \bra{B^j\tilde B^l}P_{1}\ket{A^j\tilde A^l},
    \label{ebc-6.5-13}
\end{eqnarray}
where $\ket{A^j\tilde A^l}=\ket{A^j}\otimes\ket{\tilde A^l}$, the
state $\ket{\tilde A^j}$ is the time reversed of $\ket{A^j}$,
i.e., $\tilde A^j_{m}=(-1)^m A^{j*}_{-m}$ (and similarly for
$\ket{B^j\tilde B^l}$ and $\ket{\tilde B^l}$) and $P_{1}$ is the
projector over the Hilbert space of the representation of total
spin $J=1$. Our aim is to compute
\begin{equation}
\displaystyle\aver{t}_{\rm max}=\max_{AB}\aver{t},
\end{equation}
where the maximization is over all $A^j_{m}$ and $B^j_{m}$
subject to the normalization conditions in~(\ref{ebc-6.5-1})
and~(\ref{ebc-6.5-4}). The Schwarz inequality  implies
\begin{equation}
\bra{B^j\tilde B^l}P_{1}\ket{A^j\tilde A^l}\le
\|P_{1}\ket{A^j\tilde A^l}\|\|P_{1}\ket{B^j\tilde B^l}\|,
\end{equation}
where
the equality holds iff
\begin{equation}
P_{1}\ket{A^j\tilde A^l}=\mu^{jl}\; P_{1}\ket{B^j\tilde B^l}
\qquad \forall j,l .\label{ebc-28.5-1}
\end{equation}
Hence, to compute $\aver{t}_{\rm max}$, we can restrict ourselves
to a smaller parameter space, where $\ket{A^j}$ and $\ket{B^j}$
are constrained through~(\ref{ebc-28.5-1}). This is equivalent to
consider the states  $\ket{A}$  such that
\begin{equation}
    A^j_{m}=C^j B^j_{m},\qquad\mbox{with}\quad \sum_{j}|C^j|^2=1,
    \label{ebc-6.5-18}
\end{equation}
i.e., we only need to consider the set of parameters
$\{C^j,B^j_m\}$. This we can prove, e.g., by induction on $j$
using~(\ref{ebc-28.5-1}) with $l=j+1$ and starting with the
trivial case $j=0$~\cite{escape}. Eq.~\ref{ebc-6.5-18} is easy to
understand from the physical point of view. It just tells us
that, for an optimal communication, the messenger states
$\ket{A(g)}$ must be as similar as possible to the states
$\ket{B(g)}$ on which the measuring device projects~\cite{us}. We
next substitute back in~(\ref{ebc-6.5-13}) to obtain
\begin{equation}
    \aver{t}_{\rm max}=\max_{BC}\sum_{jj'}C^j\, {\mathsf M}^{jj'}_{B}\,
    C^{j'},
    \label{ebc-6.5-19}
\end{equation}
where
\begin{equation}
    {\mathsf M}^{jj'}_{B}={\sqrt{(2j+1)(2j'+1)}\over3}
    \bra{B^j\tilde B^{j'}}P_{1}\ket{B^j\tilde B^{j'}}
    \label{ebc-6.5-20}
\end{equation}
and the maximization is over all $B^j_{m}$ and $C^j$ subject to
the normalizations~(\ref{ebc-6.5-4}) and~(\ref{ebc-6.5-18}).

Let us now discuss some properties of the matrix ${\mathsf
M}_{B}$ defined by~(\ref{ebc-6.5-20}). We first note that
${\mathsf M}_{B}$ is tridiagonal, i.e., ${\mathsf M}^{jj'}_{B}=0$
if $|j-j'|>1$, and symmetric. It is manifestly non-negative,
i.e., ${\mathsf M}^{jj'}_{B}\ge 0$ for all $j$, $j'$ and, most
important, it is rotationally invariant: any  reference state of the
form $\ket{B'}=U(g)\ket{B}$ is equally as good as
$\ket{B}$.

We next compute bounds for the diagonal ($\mathsf{M}_{B}^{jj}$)
and off diagonal ($\mathsf{M}_{B}^{j\,j+1}$) entries  of
$\mathsf{M}_{B}$. We have
\begin{eqnarray}
    &&0\le \mathsf{M}_{B}^{jj}= 
    {2j+1\over3}|\langle B^j\tilde B^j|10\rangle|^2 \le {2j+1\over3}
    \nonumber\\
     &&\times\left|\sum_{m'}|B^j_{m'}|^2\max_{m} \,\langle
     jmj-m|10\rangle\right|^2={j\over j+1},
    \label{ebc-24.5-1}
\end{eqnarray}
where we have used rotational invariance to orient the (real)
vector $P_{1}\ket{B^j\tilde B^{j}}$ along the $z$ ($m=0$) axes.
As for the off diagonal entries, the Schwarz inequality leads to
\begin{eqnarray}
     &  & 0\le \mathsf{M}_{B}^{j\,j+1}\le{\sqrt{(2j+1)(2j+3)}\over3}
     \sum_{m'}|B^j_{m'}|^2
     \nonumber\\
     &&\times
     \sum_{m''}|\tilde{B}^{j+1}_{m''}|^2
    \max_{m}\left( \sum_{M}\langle j\;M-m\;j+1\;m|1M\rangle^2\right)
     \nonumber\\
     &  & =\sqrt{2j+1\over2j+3},
    \label{ebc-24.5-2}
\end{eqnarray}
where, actually, the sum over $M$ in the second line is
independent of $m$. It is straightforward to verify that  the
particular choice
\begin{equation}
    \ket{B_{\rm op}}=\sum_{j}\sqrt{2j+1}\ket{j,j}\quad\Leftrightarrow\quad
    B^j_{{\rm op}\,m}=\de^j_{m}
    \label{ebc-6.5-21}
\end{equation}
saturates the two upper bounds~(\ref{ebc-24.5-1})
and~(\ref{ebc-24.5-2}) {\em simultaneously}. Hence
\begin{equation}
|\mathsf M_{B}^{jj'}|\le
\mathsf{M}_{B_{\rm op}}^{jj'}\equiv \mathsf{M}_{\rm op}^{jj'} ,
\label{ebc-24.5-3}
\end{equation}
for
all $j$, $j'$ and $\ket{B}$. The matrix $\mathsf{M}_{\rm op}^{jj'}$ is
\begin{equation}\label{matrix}
\mathsf{M_{\rm op}}=\pmatrix{{J\over J+1}&\sqrt{2J-1\over2J+1}&
         &        &       \cr
                    \sqrt{2J-1\over2J+1}&\ddots&\ddots
    &\phantom{\ddots} \raisebox{2.0ex}[1.5ex][0ex]{\LARGE
0}\hspace{-.5cm}       &       \cr
                         &\ddots&{2\over3}    &\sqrt{{3\over5}} &       \cr
                         & \phantom{\ddots}
                         &\sqrt{{3\over5}}&{1\over2}&\sqrt{{1\over3}}
\cr \hspace{.5cm} \raisebox{2.0ex}[1.5ex][0ex]{\LARGE
0}\hspace{-.5cm} &&\phantom{\ddots}&\sqrt{{1\over3}}&0} ,
\label{ebc-24.5-5}
\end{equation}
where $J=N/2$ is the maximum spin of the system.

We now go back to~(\ref{ebc-6.5-19}) and compute $\aver{t}_{\rm
max}$. We first note that, $\displaystyle\aver{t}_{\rm
max}=\max_{B}\la(B)$, where $\la(B)$ is the maximal eigenvalue of
the matrix ${\mathsf M}_{B}$. Since it is non-negative,
Eq.~\ref{ebc-24.5-3} implies~\cite{marcus}
\begin{equation}
    \aver{t}_{\rm max}=\max_{B}\la(B)= \la(B_{\rm op})\equiv\la_{\rm
    op}.
    \label{ebc-24.5-4}
\end{equation}
We thus have simplified the problem to that of computing
$\la_{\rm op}$, the maximal eigenvalues of $\mathsf M_{\rm op}$
in~(\ref{ebc-24.5-5}). This can be done proceeding along the same
lines as in~\cite{us,us2}. We would like to emphasize that the
calculation relies on the fact that the maximal value of each
entry of $\mathsf M_B$  is reached simultaneously, e.g.  for the
single state $\ket{B_{\rm op}}$. This is, a priori, a rather
unexpected property which, however, provides a remarkable
simplification of the calculation.

The result obtained and the form of the optimal state,
$\ket{B_{\rm op}}$, agree with our physical intuition as we now
briefly discuss. If Alice's state has a well defined total spin
(i.e. it is an eigenstate of $\vec J\,{}^2$), $\mathsf{M}_{\rm
op}$ becomes diagonal and $\aver{t}_{\rm max}=J/(J+1)=N/(N+2)$.
In terms of the average error, $\aver{h}=4(N+3)/(N+2)$, thus, at
most ($N\to\infty$) $\aver{h}=4$. In average, Bob cannot
determine more than just one axes of Alice's trihedron. The
structure of the state $\ket{B_{\rm op}}$ is  such that,  within
each irreducible representation, the determination of a single
axes is optimal~\cite{mp} (this is {\em the best} Alice could do
if she only was allowed to use a single irreducible
representation). At the same time, $\ket{B_{\rm op}}$ is as
different of an eigenstate of $J_{z}$ as it can possibly be (if
$J_z\ket{B_{\rm op}} \propto \ket{B_{\rm op}}$, Alice would be
able to communicate {\em only} a single axes).

For small $N$, one can easily obtain analytic expressions for
$\aver{t}_{\rm max}$ (see table). For large $N$ it suffices to
give simple lower and upper bounds for $\aver{t}_{\rm max}$. A
useful upper bound is provided by the condition $\aver{t}_{\rm
max}\leq \max_{j}\sum_{j'}
    {\mathsf M}_{\rm op}^{jj'}$. A lower bound is obtained
    computing $\Delta=\sum_{jj'}{C}^{j}{\mathsf M}_{\rm op}^{jj'}{C}^{j'}$
    for any normalized vector with components $C^j$. A judicious choice
    is $C^j\propto \sqrt{2j-1}(N/2-j)j^p$. The maximum of $\Delta$
    occurs at $p\approx \sqrt[3]{3N/4}$. We  obtain
\begin{equation}
  3-{4 \over N}+O(N^{-4/3})\lessapprox \aver{t}_{\rm max}\lessapprox 3-{4\over
    N}+O(N^{-2}).
    \label{ebc-6.5-24}
\end{equation}
It is now clear that perfect determination of the trihedron,
$\aver{t}_{\rm max}=3$, is reached in the asymptotic limit, and
$\aver{t}$ approaches three at most linearly in $1/N$. Finally, we
have performed a linear fit obtaining
\begin{equation}\label{fit}
  \aver{t}_{\rm max} \sim 3 -\frac{4}{N}-\frac{9.4}{N^{4/3}}+\dots,
\end{equation}
which is completely consistent with (\ref{ebc-6.5-24}).
%
\begin{table}\label{table-I}
\begin{center}
\begin{tabular}{c|cccccc}
  \toprule
  $N$  & 2 & 3 & 5 & 10 & 50 & 100 \\
  \colrule\phantom{\big|}
  $\aver{t}_{\rm max} $ & $\frac{3+\sqrt{57}}{12}$ &
  $\frac{14+\sqrt{466}}{30}$  & 1.6708  & 2.6202 & 2.9362 &  2.9707 \\
  \botrule
\end{tabular}
\end{center}
\caption{Maximal value of $\langle t\rangle$ vs. the number of spins}
\end{table}

We now turn our attention to the construction of POVM's with a
finite number of outcomes, as they are the only ones that can be
physically realized.  The main idea is stated in \cite{us3}.
There, we introduced the concept of set of directions
isotropically distributed. In the context of the present letter
the term directions has to be generalized  to elements of the
group. We say that a finite set $\{g_r\}$, $r=1,\cdots N(J)$, of
elements of $\mathrm{SU(2)}$ is isotropically distributed up to
spin $J$, if there exist positive weights $\{c_{r}\}$ such that
the following orthogonality relation holds for any $j,j'\leq J$:
\begin{equation}
\sum_{r=1}^{N(J)}  c_{r}\,\D{j}{m}{n} (g_{r}) \Dc{j'}{m'}{n'}
(g_{r})
 =\frac{C_J}{2j+1}\delta_m^{m'}\delta_n^{n'}\delta_{j}^{j'},
 \label{discrete-1}
\end{equation}
where $C_J=\sum_{r=1}^{N(J)} c_r$. This discrete version of
(\ref{ebc-6.5-5}) is only valid up to a certain value $J$,  the
larger $J$ is, the larger $N(J)$ must be chosen. Working along
the same lines as in \cite{us3} one can show  that the angular
dependence on $\alpha$ and $\gamma$ can be trivially satisfied
choosing $N+1$ equidistant angles for each variable. The only
non-trivial conditions concern the set $\{\beta_r\}$, which is
required to satisfy $ \label{eqzz} \sum_{r} c_{r}
P_{L}(\cos{\beta_{r}})=0$ ($1\leq L\leq 2J$), where $P_{L}$ is
the Legendre polynomial of degree~$L$. The procedure to solve
this  equation is described in  \cite{us3} (see also \cite{pp2}).
This recipe yields a finite optimal POVM for any value of~$N$. In
general, however, one can find  equally optimal POVM's with a
smaller number of outcomes. Ideally one would be interested in
finding the minimal ones, however, as far as we are aware, the
solution is not known for arbitrary $J$ and general
groups~\cite{lpt}.

Nevertheless, the minimal POVM for the first non-trivial case of
two spins  is not difficult to find. Consider the simplest
normalized reference state that leads to an optimal POVM
\begin{equation}\label{B-2}
\ket{B}=\frac{\sqrt{3}}{2}\ket{1,1}+\frac{1}{2}\ket{0,0},
\end{equation}
It is easy to verify that the four projectors
$O_r=U(g_r)\ket{B}\bra{B}U^\dagger(g_r)$, with
\begin{equation}
\begin{array}{lllr}
\alpha_r=(r-1)\frac{2\pi}{3},&
\gamma_r=\pi-\alpha_r,&\cos\theta_r=\frac{1}{3},& r\le3\\
\alpha_4=0,&\gamma_4=0,&\cos\theta_4=1,&
\end{array}
\label{tetrahedron}
\end{equation}
satisfy the POVM condition $\sum_{r=1}^{4}O_r=\mathbb I$. Since
the Hilbert space has dimension four, the minimal number
of outcomes for any measurement is also four. This measurement is therefore
finite, minimal, and optimal. In fact, it is a von Neumann
measurement as $O_r O_s=\delta_{rs}O_r$. Notice that the set of
points (\ref{tetrahedron}) do not satisfy the orthogonality
conditions (\ref{discrete-1}) for all the $m,m'$ values, but it
does for the relevant ones. It is the particular structure of the
state (\ref{B-2}) what enables us to construct a POVM with only
four outcomes.

We conclude that it is feasible to use quantum systems to encode
the orientation of a reference frame. The optimal strategy
involves the use of encoding states which are remarkably simple
and have a clear physical interpretation. The average error of the
transmission is seen to approach zero linearly in $1/N$. Finally,
we give a recipe for constructing finite optimal POVMs and present
an example of a minimal one for the simple case $N=2$.

Financial support from CICYT contract AEN99-0766 and CIRIT
contracts 1998SGR-00051, 1999SGR-00097 is acknowledged.


\begin{thebibliography}{99}
\bibitem{escape} One can try to escape from~(\ref{ebc-6.5-18}) by further
constraining $\ket{B^j}$ in such a way that the orthogonal complement
of
$P_{j+1}({\cal H}^1\otimes\ket{B^j})$, where ${\cal H}^1$  is
the Hilbert space of the $j=1$ representation of $\rm SU(2)$ and
$P_{j+1}$ is the projector over ${\cal H}^{j+1}$, is different from zero.
Obviously, this leads to non optimal states $\ket{A^j}$ and $\ket{B^j}$.
\end{thebibliography}
\end{document}